\documentclass[pra,twocolumn,showpacs]{revtex4}
\usepackage{graphicx}
\usepackage{amsmath}
\usepackage{amssymb}
\newcommand{\sss}{\scriptstyle}
\def\lsim{\
  \lower-1.2pt\vbox{\hbox{\rlap{$<$}\lower5pt\vbox{\hbox{$\sim$}}}}\ }
\def\gsim{\
  \lower-1.2pt\vbox{\hbox{\rlap{$>$}\lower5pt\vbox{\hbox{$\sim$}}}}\ }

\begin{document}

\title{THEORY OF A NARROW ROTON ABSORPTION LINE \\
    IN THE SPECTRUM OF A DISK-SHAPED SHF RESONATOR}

\author{Vadim M.~Loktev}
\email{vloktev@bitp.kiev.ua}
\author{Maksim D.~Tomchenko}
\email{mtomchenko@bitp.kiev.ua}
\affiliation{Bogoliubov Institute
for Theoretical Physics,
        14-b Metrologichna Street, Kyiv 03680, Ukraine}

\date{\today}
\begin{abstract}
We calculate the probability of the creation of a circular phonon
(c-phonon) in
        He II by a c-photon of the resonator. It is shown that this
        probability has sharp maxima at frequencies, where
        the effective group velocity of the c-phonon is equal to
        zero; the density of states of c-phonons strongly grows
         at such frequencies. For He II, these frequencies
        correspond to a roton and a maxon. From the
        probability of the c-roton creation, we calculate the roton
        line width which is found to approximately agree with the experimental one.
        We conclude that the roton line observed in the super-high-frequency (SHF)
        absorption spectrum of helium is related to the creation
         of c-rotons. A possible interpretation
       of the Stark effect observed for the roton line
              is also proposed.
       \end{abstract}


\maketitle

  \section{Inroduction}
          In the recent works \cite{svh1,svh3}, an unconventional effect was discovered. In a dielectric disk-shaped
     resonator placed in liquid $^4$He,  the azimuth ($l$-)modes were excited. These modes are of the ``whispering-gallery'' type
     and represent a superposition of standing and running
     resonance electromagnetic (EM)  waves of the  SHF range in a narrow  frequency interval. At
       the roton frequency  $\nu_{rot} = \Delta_{rot}/2\pi\hbar =
     180.3\,\mbox{GHz},$
       the supernarrow absorption line with the width ${ \sss\triangle} \nu \simeq 50\,\mbox{kHz}$ was observed in the SHF spectrum of the resonator.
      This width is by six orders less than that of a roton peak measured in neutron experiments
      and is comparable with the width of line in the M\"{o}ssbauer effect.
      It was assumed in \cite{svh3} that the line is related to the Van Hove singularity caused by a
      high density of states of plane (p-) rotons near the roton minimum of the dispersion curve.
      In this case, however, one is faced with the problem to satisfy the momentum conservation law,
      since the p-roton momentum is greater than the momentum of a p-photon with the same energy by six orders.
      Therefore, it was supposed \cite{svh3} that the excess momentum of a p-roton is transferred to helium as a whole.

         To clarify this and other points, it is necessary to calculate the probability of the creation of a roton and the widths
         and the forms of lines
         for various possible processes, and then to choose a process explaining the experiment.
         It is necessary to take into account that the EM field of a disk-shaped resonator
         has the circular (c-) symmetry and is concentrated only near the disk according to measurements \cite{svh1} and the theory \cite{I}.
         Since namely the EM field induces the transition, the latter must be characterized by the c-symmetry.
         In \cite{I}, it was shown that a phonon near a disk possesses also the c-symmetry. Therefore, we assume that
          the narrow line corresponds to the creation of a c-roton  by the EM field of the resonator.
          Below, we will find the probability of this process and the width of the corresponding absorption line.

         A number of results required for calculations was obtained in \cite{I};
         formulas (N) from that work will be denoted here by (N$^*$). A part of the results of the present work was briefly published
         in \cite{II}.

             \section{Probability of the creation of a circular phonon by the field of a resonator}

             To calculate the line width, it is necessary to know the probability of the c-photon
             $\rightarrow$ c-phonon process. No problems concerning the conservation laws appear for this process, because
             both c-photon and c-phonon have no momentum, but have angular momenta $L_{z}=\hbar l$
             and $L_{z}=\hbar l_{c}$, respectively \cite{I}. Moreover, the condition $ l=l_{c}$ is easily satisfied.
             As will be seen, the field of the resonator contains $\sim 10^{12}$ photons with close frequencies.
             At such occupation number, the photon field can be considered as an
             external classical perturbing harmonic field with frequency $\nu$ acting on
             helium. We now calculate the probability of the creation of a c-phonon
             by this field. The width of the azimuth mode, on which
             the roton line is observed, is about 2.5\,MHz, the line width $\sim 0.1\,$MHz, and the frequency
             $\nu=180.3$ GHz. It is seen that the mode width is very small as compared with
             the frequency; therefore, the latter can be considered constant.

             It is worth noting that, while calculating the transition amplitude, we can take no care of the conservation laws.
    They are satisfied automatically (if a transition contradicts some
    conservation law, this will manifest itself in the disagreement of the symmetries of the
    initial and final states, and the amplitude will become
    equal to zero).

             The probability (per unit time) of the creation of a c-phonon in He II
             due to the action of the EM field of a resonator is \cite{land3}
               \begin{equation}
             \delta w_{fi} = \frac{2\pi}{\hbar}|F_{fi}|^{2}\delta (E_{f}-E_{i}^{(0)}-\hbar\omega),
           \label{a1} \end{equation}
                         \begin{equation}
             F_{fi} = \int\Psi_{f}\hat{F}\Psi_{i}d\Omega^{nuc}d\Omega^{el},
           \label{a2} \end{equation}
              where $d\Omega^{nuc}=d\textbf{R}_{1}\ldots d\textbf{R}_{N}$ and
               $d\Omega^{el}=d\textbf{R}_{1}^{(1)}d\textbf{R}_{1}^{(2)}\ldots
               d\textbf{R}_{N}^{(1)}d\textbf{R}_{N}^{(2)}$ are the
               phase volumes of all nuclei and all
               electrons, $\Psi_{i}$ and $\Psi_{f}$ are the wave functions (WFs) of the initial and final states of helium, respectively,
             $E_{f}-E_{i}^{(0)}=E_{c}$ is the energy of a
             c-phonon, and  $\omega = 2\pi\nu$. The explicit formula for $\hat{F}$ follows from the perturbation operator
            \begin{equation}
             \hat{V}  = \hat{F}e^{-i\omega t} + \hat{F}^{+}e^{i\omega t}.
           \label{a3} \end{equation}
                                 If the wavelength $\lambda$ of the EM field is much more than the size of
             the system, then the problem is solved in the dipole approximation
             \cite{land4}. In our case, this approximation is not suitable, since the system size exceeds $\lambda$ by one order.
             In addition, a photon
             is spent on the excitation of fluid helium as a whole, i.e. on the creation of a c-phonon which is related to the motion of atoms as united objects,
             rather than on the excitation of electron shells of a single atom or many atoms.
             Therefore, we will use a general approach,
             by considering the action of the EM field directly on the charged particles in an atom, i.e., on electrons and the nucleus.
             But the atoms interact with one another. As a result,
              the EM field creates
               the collective excitation, a phonon which is electrically neutral as a whole. It should be noted that a sound
                wave is associated with a variable concentration gradient. In this case, a variable local electric field arises
             in the interatomic space, since helium atoms polarize one another
             \cite{wb,lt2}. However, this gives only a negligible correction to the effect.
             For a charge particle in the EM field, we have \cite{land3}
             \begin{eqnarray}
             \hat{V}  &=& -\frac{q}{2mc}(\textbf{A}\hat{\textbf{p}}
             + \hat{\textbf{p}}\textbf{A}) +
             \frac{q^2}{2mc^2}\textbf{A}^2 = \nonumber \\
             &=& -\frac{q}{mc}\left (\textbf{A}\hat{\textbf{p}}
             - \frac{i\hbar}{2}div\textbf{A}\right ) +
             \frac{q^{2}\textbf{A}^2}{2mc^2},
           \label{a4} \end{eqnarray}
               where $\hat{\textbf{p}}=-i\hbar\nabla_{\textbf{r}}$, and $q$ and $\textbf{r}$ are the charge and the radius-vector of a particle, respectively.
               The term with $\textbf{A}^2$ induces two-photon transitions which will not be considered here.
               The field $\textbf{A}$ should be real and can be presented in the form
               \begin{equation}
             \textbf{A}(\textbf{r},t)  = \textbf{A}_{0}(\textbf{r})e^{-i\omega t} + \textbf{A}_{0}^{*}(\textbf{r})e^{i\omega t}
             + \tilde{\textbf{A}}(\textbf{r}).
           \label{a5} \end{equation}
           The quantity $div\textbf{A}$ is calculated in \cite{I}, formula (6$^*$). Since
           we are interesting in the field $\textbf{A}$ in helium,
            we set $\varepsilon _{\bot} = \varepsilon
            _{z} \equiv \varepsilon _{h}$ in (6$^*$). As a result, we obtain
            $div\textbf{A}=f_{d}(\textbf{r})$, i.e. the divergence is
            determined by the field $\tilde{\textbf{A}}(\textbf{r})$, i.e. the
            time-independent part of $\textbf{A}$. It can be always set
            to zero, by adding a gradient of the corresponding function to $\textbf{A}.$
            In this case,
            the measurable quantities $\textbf{E}$ and  $\textbf{H}$ are
            not changed. Therefore, we set $\tilde{\textbf{A}}(\textbf{r})=0$, so that
            $div\textbf{A}=0$.
              In what follows, we omit index 0 in $\textbf{A}_{0}$ (\ref{a5}). So, we have
                \begin{equation}
             \hat{F}  = \frac{i\hbar q}{mc}\textbf{A}\nabla_{\textbf{r}}.
                \label{a6}  \end{equation}
                Consider the set of He II atoms in the field
                $\textbf{A}$. An atom of $^4$He consists of the nucleus with the charge $q_{n}=-2e$
                and two electrons, each possessing the charge $q_{e}=e$.
                Therefore, for He II, we have
                 \begin{eqnarray}
          \hat{F} & = &\sum\limits_{j}\frac{i\hbar}{c}\left ( \frac{q_{n}}{m_{n}}\textbf{A}(\textbf{R}_{j})\frac{\partial}{\partial\textbf{R}_{j}}
           + \frac{q_{e}}{m_{e}}\textbf{A}(\textbf{R}^{(1)}_{j})\frac{\partial}{\partial\textbf{R}^{(1)}_{j}} \right. \nonumber \\
           &+& \left.
           \frac{q_{e}}{m_{e}}\textbf{A}(\textbf{R}^{(2)}_{j})\frac{\partial}{\partial\textbf{R}^{(2)}_{j}}\right ).
                \label{a7}  \end{eqnarray}
                        Here, $\textbf{R}_{j}$ are coordinates of the nucleus of the $j$-th atom,
             $\textbf{R}^{(1)}_{j}$ and $\textbf{R}^{(2)}_{j}$ are coordinates of the electrons of this atom,
              $m_{n}\approx m_{4}$ is the nucleus mass
             of $^4$He atom, and $m_{e}$ is the electron mass.

                To calculate $F_{fi},$ we need to know the WFs $\Psi_{f}$ and $\Psi_{i}$.
                Let the initial state characterized by the WF $\Psi_{i}$ be the ground
                state of helium, and let the final state with $\Psi_{f}$ be the ground
                state plus one c-phonon.  Then
                 \begin{equation}
            \Psi_{i} \equiv \Psi_{0}(\{\textbf{R}_{j},\textbf{R}^{(1,2)}_{j}\})
             = \Psi_{0}^{nuc}(\{\textbf{R}_{j}\})\Psi_{0}^{el}(\{\textbf{R}^{(1,2)}_{j}\}),
                \label{a8}  \end{equation}
                          where $\Psi_{0}^{el}$ is the WF of all electrons, and
            $\Psi_{0}^{nuc}$ is the WF of all nuclei of helium atoms.
            The modern microscopic models describe the properties of He II well enough.
            According to them, a helium atom is a very elastic object. Therefore, we can consider with a good accuracy that
            the electron shell of each atom follows the nucleus without inertia and is not
            deformed by the interaction with neighboring atoms. In this
            case, $\Psi_{0}^{nuc}(\textbf{R}_{j})$ coincides with
             $\Psi_{0}(\textbf{R}_{j})$  written as a function of the coordinates of atoms,
            and the electron part looks as
             \begin{eqnarray}
            \Psi_{0}^{el} &=& \prod\limits_{j=1}^{N}\psi_{j}(\textbf{R}_{j}^{(1)},\textbf{R}_{j}^{(2)})\approx
            \prod\limits_{j=1}^{N}\psi_{1s^2}(\textbf{r}_{j}^{(1)},\textbf{r}_{j}^{(2)}) \approx \nonumber \\
         &\approx & \prod\limits_{j=1}^{N}\tilde{\psi}_{1s}(\textbf{r}_{j}^{(1)})\tilde{\psi}_{1s}(\textbf{r}_{j}^{(2)}),
                \label{a13}  \end{eqnarray}
             where  $\textbf{r}^{(1)}_{j}=\textbf{R}_{j}^{(1)}-\textbf{R}_{j}$,
              $\textbf{r}^{(2)}_{j}=\textbf{R}_{j}^{(2)}-\textbf{R}_{j}$.
              For the WF of the ground state of $^4$He atom, we use the well-known one-parameter
              approximation
              with $\tilde{\psi}_{1s}(\textbf{r})=\frac{1}{\sqrt{\pi
              a^3}}e^{-r/a}$ ($a=0.313 \mbox{\AA}$). In reality, the electron
              shells of atoms are perturbed by neighbors and as a result are somewhat
              deformed \cite{wb,lt2}. These deformations are
              ``directed'' to the atom, with which the interaction occurs.
              Since the adjacent atoms surround the given atom from all sides and
              chaotically on the average, the polarizations induced by them cancel one another to a significant
              degree. As a result, the mean polarization of an atom in He II turns out small,
              $d \simeq 3\cdot 10^{-4} |e| a_{B}$ \cite{pol2D}, though it is greater by one order of magnitude than the polarization induced by a single atom
              located at the mean interatomic distance
              $ \bar{R} \approx 3.6\,\mbox{\AA}$. The literature presents the assertions that the presence of
              large-amplitude zero oscillations of He II with the amplitude $\sim \bar{R}$
               implies that the electron shell of a helium atom is strongly deformed and ``smeared'' over several interatomic distances.
              However, this should not be accepted: zero oscillations mean only that, due to the strong interatomic interaction, the helium atoms
               move with a large (even at $T=0$) mean velocity, which leads to a small density of helium and
               exhausts the condensate. But the electron shells of atoms are deformed slightly in this case, as follows from the smallness of
               $d$. We will neglect these deformations and apply formula (\ref{a13}).

                     For the WF of a state with a single
              c-phonon, we use the formula \cite{I}
               \begin{equation}
            \Psi_{f} = \Psi_{0}\Psi_{c}(l_{c},k_z,k_{\rho}).
                \label{a14}  \end{equation}
             In view of it, we obtain
         \begin{eqnarray}
             F_{fi} &\approx & \frac{2i\hbar |e|N}{cm_{4}}   \int\Psi_{0}\Psi_{c}^{*}
             \left ( \textbf{A}(\textbf{R}_{1})\frac{\partial}{\partial\textbf{R}_{1}}\right. - \nonumber \\
           &-& \left. \frac{m_{4}}{m_{e}}\textbf{A}(\textbf{R}^{(1)}_{1})\frac{\partial}{\partial\textbf{R}^{(1)}_{1}}
           \right )      \Psi_{0}d\Omega^{nuc}d\Omega^{el}.
           \label{a15} \end{eqnarray}
           Let us take into account that
           \begin{equation}
            \frac{\partial \Psi_{0}}{\partial\textbf{R}_{1}} = \Psi_{0}^{el}\frac{\partial \Psi_{0}^{nuc}}{\partial\textbf{R}_{1}}
            +\Psi_{0}^{nuc}\frac{\partial \Psi_{0}^{el}}{\partial\textbf{R}_{1}},
                \label{a16}  \end{equation}
            \begin{equation}
            \frac{\partial \Psi_{0}^{el}}{\partial\textbf{R}_{1}^{(1)}} =
             \frac{\partial \tilde{\psi}_{1s}(\textbf{r}_{1}^{(1)})}{\partial\textbf{r}_{1}^{(1)}}
             \frac{\Psi_{0}^{el}}{\tilde{\psi}_{1s}(\textbf{r}_{1}^{(1)})},
                \label{a17}  \end{equation}
          \begin{equation}
          \frac{\partial \Psi_{0}^{el}}{\partial\textbf{R}_{1}} =
          -\left [\frac{\partial \tilde{\psi}_{1s}(\textbf{r}_{1}^{(1)})}{\partial\textbf{r}_{1}^{(1)}}
             \frac{1}{\tilde{\psi}_{1s}(\textbf{r}_{1}^{(1)})} + (\textbf{r}_{1}^{(1)} \leftrightarrow
             \textbf{r}_{1}^{(2)}) \right ]\Psi_{0}^{el}.
           %  \frac{\partial \tilde{\psi}_{1s}(\textbf{r}_{1}^{(2)})}{\partial\textbf{r}_{1}^{(2)}}
           %  \frac{1}{\tilde{\psi}_{1s}(\textbf{r}_{1}^{(2)})}\right ]\Psi_{0}^{el}.
                \label{a18}  \end{equation}
    Then
        \begin{equation}
            F_{fi} = F_{fi}^{nuc} + F_{fi}^{el},
                \label{a19}  \end{equation}
        \begin{equation}
            F_{fi}^{nuc} = \frac{i\hbar |e|N}{m_{4}c}   \int\Psi_{c}^{*}
            \textbf{A}(\textbf{R}_{1})\frac{\partial}{\partial\textbf{R}_{1}}
            (\Psi_{0}^{nuc})^{2}d\Omega^{nuc},
                \label{a20}  \end{equation}
         \begin{eqnarray}
           & F_{fi}^{el} &= -\frac{2i\hbar |e|N}{m_{4}c}   \int\Psi_{c}^{*}
            (\Psi_{0}^{el}\Psi_{0}^{nuc})^{2}
            \frac{\partial \tilde{\psi}_{1s}(\textbf{r}_{1}^{(1)})}{\partial\textbf{r}_{1}^{(1)}}
             \frac{1}{\tilde{\psi}_{1s}(\textbf{r}_{1}^{(1)})}  \nonumber \\
             &\times & \left
             (\textbf{A}(\textbf{R}_{1}+\textbf{r}_{1}^{(1)})\frac{m_{4}}{m_{e}}
             +2\textbf{A}(\textbf{R}_{1}) \right )d\Omega^{nuc}d\Omega^{el}.
                \label{a21}  \end{eqnarray}
     \subsection{Calculation of $F_{fi}^{nuc}$} \label{2.1}
         We note that the matrix element $F_{fi}$ in (\ref{a1}) is constructed on the basis of the WFs
    $\Psi_{i}$ and $\Psi_{f}$ not containing the time factor (the latter is included into the $\delta$-function).
       In formula (50$^*$) for $\Psi_{c},$ the coordinates of atoms can be considered as the coordinates of nuclei.
       Omitting the factor  $e^{-i\omega t}$, we present $\Psi_{c}$  in the form
         \begin{equation}
  \Psi_{c} \approx \sum\limits_{j=1}^N \Psi_{c}(j), \
  \Psi_{c}(j)   =  \frac{c_{l_{c},k_z,k_{\rho}}}{\sqrt{N}} e^{il_{c}\varphi_{j}+ik_{z}Z_{j}}
  J_{l_{c}}(k_{\rho}\rho_{j}),
            \label{a22} \end{equation}
  below $c_{l_{c},k_z,k_{\rho}} \equiv  \tilde{c}$.   Whence, with regard for (56$^*$) and (58$^*$), we get
   \begin{eqnarray}
     & F_{fi}^{nuc}& = \frac{i\hbar |e|N}{m_{4}c} \left \{  \int d\textbf{R}_{1}
      \Psi_{c}^{*}(1)\textbf{A}(\textbf{R}_{1})\frac{\partial}{\partial\textbf{R}_{1}}
    \frac{1}{V}  + \right. \label{a23}  \\
    &+& \left. (N-1)\int d\textbf{R}_{1}d\textbf{R}_{2}
      \Psi_{c}^{*}(2)\textbf{A}(\textbf{R}_{1})\frac{\partial g(|\textbf{R}_{1} - \textbf{R}_{2}|)}{V^{2} \partial\textbf{R}_{1}}
        \right \}.
             \nonumber     \end{eqnarray}
    Since $\frac{\partial}{\partial\textbf{R}_{1}}
    \frac{1}{V}=0$, the first integral in (\ref{a23}) is equal to zero. As for the second
    integral, the integration is carried on over the region, where the helium atoms are positioned, i.e., over the region outside of the disk.
    Therefore, we describe the field $\textbf{A}$ with the use of solution (31$^*$)--(33$^*$)
    without the factor $e^{-i\omega t}$
    (taken into account in (\ref{a5})). Denoting $|\textbf{R}_{1} - \textbf{R}_{2}|=R$,
      we have
    \begin{eqnarray}
  &\textbf{A}(\textbf{R}_{1})\frac{\partial g(R)}{\partial\textbf{R}_{1}}&  =
         A_{m}e^{il\varphi_{1}} \left [a_{1}(\rho_{1},Z_{1})
        \left (\frac{\partial}{\partial\rho_{1}}+
        \frac{i}{\rho_{1}}\frac{\partial}{\partial\varphi_{1}} \right )\right. + \nonumber \\
      &+&  \left. a_{2}(\rho_{1},Z_{1})
        \left (-\frac{\partial}{\partial\rho_{1}}+
        \frac{i}{\rho_{1}}\frac{\partial}{\partial\varphi_{1}} \right )
        \right ] g(R).
         \label{a24} \end{eqnarray}
        Using (72$^*$)  and the relation $\textbf{q}\textbf{R}_{1}=
        q_{z}Z_{1} +q_{\rho}\rho_{1}\cos{(\varphi_{1}-\varphi_{q})}$, we obtain
   \begin{eqnarray}
  &&\textbf{A}(\textbf{R}_{1})\frac{\partial g(R)}{\partial\textbf{R}_{1}}
  =  \frac{A_{m}e^{il\varphi_{1}}}{(2\pi)^{3}n} \int
  d\textbf{q}[S(q)-1]iq_{\rho}\times \nonumber \\
  && \times \left [a_{1}(\rho_{1},Z_{1})e^{-i(\varphi_{1}-\varphi_{q})}
  - a_{2}(\rho_{1},Z_{1})e^{i(\varphi_{1}-\varphi_{q})} \right ]\times
     \nonumber \\ && \times e^{[iq_{z}(Z_{1}-Z_{2}) + iq_{\rho}\rho_{1}\cos{(\varphi_{1}-\varphi_{q})}  -
         iq_{\rho}\rho_{2}\cos{(\varphi_{2}-\varphi_{q})}]},
        \label{a25}  \end{eqnarray}
         where $n=N/V$.
   In the cylindrical coordinate system, $d\textbf{R}_{j}=d\varphi_{j}dZ_{j}\rho_{j}d\rho_{j}$ and
   $d\textbf{q}=d\varphi_{q}dq_{z}q_{\rho}dq_{\rho}$.
    The relation
      \begin{equation}
          J_{l}(x) =\frac{i^l}{2\pi} \int\limits_{-\pi+\beta}^{\pi+\beta}
          e^{-ix\cos{\psi}\pm il\psi}d\psi
          \label{I64} \end{equation}
       ($\beta$ is arbitrary) yields
     \begin{eqnarray}
      && \int\limits_{0}^{2\pi}d\varphi_{1} e^{[il\varphi_{1}\pm i(\varphi_{1}-\varphi_{q})
          +iq_{\rho}\rho_{1}\cos{(\varphi_{1}-\varphi_{q})}]} = \nonumber \\
          &&= 2\pi
          i^{l\pm 1}e^{il\varphi_{q}}J_{l\pm 1}(q_{\rho}\rho_{1}),
          \label{a26} \end{eqnarray}
    \begin{eqnarray}
       &&  \int\limits_{0}^{2\pi}d\varphi_{2} \exp{[-il_{c}\varphi_{2}
          -iq_{\rho}\rho_{2}\cos{(\varphi_{2}-\varphi_{q})}]} = \nonumber \\
          && = 2\pi
          i^{-l_{c}}e^{-il_{c}\varphi_{q}}J_{l_{c}}(q_{\rho}\rho_{2}).
          \label{a27} \end{eqnarray}
  With regard for (32$^*$), (33$^*$)  and (\ref{a22}), we write the required matrix element as
  \begin{eqnarray}
     && F_{fi}^{nuc} = \frac{i\delta_{l,l_{c}}\hbar |e|n\tilde{c}A_{m}}{m_{4}c\sqrt{N}}
       \int dZ_{1}dZ_{2}\rho_{1}d\rho_{1}\rho_{2}d\rho_{2}dq_{z}\times \nonumber \\
      && \times q_{\rho}^{2}dq_{\rho}J_{l}(q_{\rho}\rho_{2})J_{l}(k_{\rho}\rho_{2})
        [S(q)-1]e^{[iq_{z}(Z_{1}-Z_{2})-ik_{z}Z_{2}]}\times \nonumber \\
       && \times\left [a_{1}(\rho_{1},Z_{1})J_{l-1}(q_{\rho}\rho_{1})+a_{2}(\rho_{1},Z_{1})J_{l+1}(q_{\rho}\rho_{1})
       \right ]. \label{a28}
                  \end{eqnarray}
       Integrating $\textbf{R}_{2}$ over the whole volume (including the disk), we obtain
        \begin{eqnarray}
        && \int dq_{z}dZ_{2}e^{[iq_{z}(Z_{1}-Z_{2})-ik_{z}Z_{2}]}
         [S(q_{z},q_{\rho})-1]= \nonumber \\
         && = 2\pi
         e^{-ik_{z}Z_{1}}[S(-k_{z},q_{\rho})-1].
          \label{a29} \end{eqnarray}
          We pass in (\ref{a28}) to discrete $q_{\rho}$, by replacing
     $\int dq_{\rho} \rightarrow \frac{\pi}{R_{\infty}-\tilde{R}_{d}}\sum
         \limits_{n_{\rho}}$ and using for $q_{\rho}$ the quantization conditions  \cite{I}
           \begin{equation}
   k_{z}=\frac{2\pi n_{z}}{H-h_{d}}, \ n_{z}=\pm 1,\pm 2,...,
    \label{I42a} \end{equation}
      \begin{equation}
    k_{\rho}=\frac{\pi n_{\rho}}{R_{\infty}-\tilde{R}_{d}}, \ n_{\rho}\gg l
   \label{I42b} \end{equation}
   where $h_{d}=0.1\,cm$ and $R_{d}=0.95\,cm$ are, respectively, the height and the radius of the disk resonator,
    and $H\approx 4.2\,cm$ and $R_{\infty}\approx 2.1\,cm$
     are the same for the chamber with helium; the numbers are indicated for the  experiment \cite{rub3}.
     The quantity $\tilde{R}_{d}$  is defined in \cite{I}: for small $k_{\rho},$ $\tilde{R}_{d}>R_{d}$
         (in particular, for the smallest $k_{\rho},$ we have $\tilde{R}_{d}\approx 1.5R_{d}$),
         and $\tilde{R}_{d}$ decreases down to $R_{d}$ with increase in $k_{\rho}$.  With regard for (64$^*$),
          we will take the integral required in the calculations of (\ref{a28}):
          \begin{eqnarray}
          I_{p}  &\equiv& \int \rho_{2} d\rho_{2}q_{\rho}^{2}dq_{\rho}[S(-k_{z},q_{\rho})-1]\times \nonumber \\
           &\times & J_{l}(q_{\rho}\rho_{2})J_{l}(k_{\rho}\rho_{2})J_{l\pm 1}(q_{\rho}\rho_{1}) \approx\nonumber \\
            &\approx &  k_{\rho}[S(k)-1]J_{l\pm 1}(k_{\rho}\rho_{1})
            \frac{B(k_{\rho}R_{\infty},l)}{1-\tilde{R}_{d}/R_{\infty}},
     \label{a31} \end{eqnarray}
            \begin{equation}
      B(x,l) = \pi x \int\limits_{0}^{1}ydy J_{l}^{2}(y\cdot x).
                \label{I57}  \end{equation}
       At greater $k_{\rho},$ the function  $B(k_{\rho}R_{\infty},l)\approx 1$, whereas it is close to 1 ($\gsim 0.9$) at lower $k_{\rho}$ ($k_{\rho}R_{\infty} \lsim l$)
       and can be calculated numerically.

                    Taking the disk into account gives $I_{p}\rightarrow I_{p}\left (
       1-\frac{h_{d}}{H}\frac{R_{d}}{R_{\infty}}\frac{B(k_{\rho}R_{d},l)}{B(k_{\rho}R_{\infty},l)} \right )$.
           Since  $h_{d}/H\simeq 1/42$,  $R_{d}/R_{\infty} \simeq 1/2$, and $B \approx 1$, this consideration changes
         $I_{p}$ only slightly, which can be neglected.
        From (\ref{a28})--(\ref{a31}), we now obtain
       \begin{equation}
      F_{fi}^{nuc} = i\delta_{l,l_{c}}\xi\sqrt{k_{\rho}^{3}}I_{nuc},
      \label{a32a}  \end{equation}
         \begin{equation}
        \xi = \frac{2\pi\hbar |e|n\tilde{c}A_{m}}{m_{4}c\sqrt{Nk_{\rho}}}
       \frac{B(k_{\rho}R_{\infty},l)}{1-\tilde{R}_{d}/R_{\infty}}[S(k)-1],
                \label{a32b}  \end{equation}
       \begin{eqnarray}
           I_{nuc}&=&  \int dZ_{1}\rho_{1} d\rho_{1}e^{-ik_{z}Z_{1}}
           \left [a_{1}(\rho_{1},Z_{1}) J_{l-1}(k_{\rho}\rho_{1})\right. + \nonumber \\
         &+&  \left. a_{2}(\rho_{1},Z_{1}) J_{l+1}(k_{\rho}\rho_{1})\right ].
     \label{a33} \end{eqnarray}
      As is seen, we must integrate with respect to the variables $Z_{1}$ and $\rho_{1}$
      over the region outside of the disk with regard for the distribution of the EM field calculated in \cite{I}.
       We denote the values of the integral $I_{c}$ taken over regions I and II from (32$^*$)--(33$^*$)  by, respectively, indices 1 and 2:
          \begin{equation}
           I_{nuc}= I_{nuc}^{(1)} + I_{nuc}^{(2)}.
     \label{a33-2} \end{equation}
      It will be seen below that values of the functions $F_{fi}^{nuc}$ and  $F_{fi}^{el}$ are maximum in two limiting cases:
      $k_{\rho} \rightarrow \infty$ and $k_{z} \rightarrow \infty$.
      In the first case, the simple, but sufficiently bulky calculations give
            \begin{equation}
           I_{nuc}^{(1)} \approx \frac{\cos{\alpha_{l}(k_{\rho}R_{d})}}{k_{\rho}^{2}tg^{2}\theta_{0}}\sqrt{\frac{2R_{d}}{\pi k_{\rho}}}
           \cdot O\left (\frac{k_{z}+\kappa_{z}+Q_{1}tg\theta_{0}}{k_{\rho}}\right ),
                             \label{a34} \end{equation}
         \begin{equation}
           I_{nuc}^{(2)} \approx -\frac{(2l-2)\sin{\alpha_{l}(k_{\rho}R_{d})}}{k_{\rho}^{2}}
           \sqrt{\frac{2}{\pi k_{\rho}R_{d}}}I_{z}(k_{z},R_{d}),
                             \label{a35} \end{equation}
        \begin{eqnarray}
      %    && I_{z}(k_{z},\rho)\equiv \int\limits_{-\breve{h}_{d}}^{\breve{h}_{d}}dZ\cdot e^{-ik_{z}Z}
      %    J_{l-1}(Q_{1}R_{d})\cos{\frac{Z\pi}{\tilde{h}_{d}}}
        I_{z}(k_{z},\rho)   & = &  \frac{J_{l-1}(Q_{1}R_{d})\sin{\left [(\pi/\tilde{h}_{d}-k_{z})\breve{h}_{d}\right ]}}{\pi/\tilde{h}_{d}-k_{z}}
           +  \nonumber \\  &+& (k_{z}\rightarrow -k_{z}),
           \label{a36} \end{eqnarray}
           where $\alpha_{l}(k_{\rho}R_{d})=k_{\rho}R_{d}-\pi l/2 - \pi/4$, $\breve{h}_{d}=
       h_{d}/2+(\rho - R_{d})ctg\theta (\rho)$, $\tilde{h}_d \approx 1.087 h_{d}$, and $\theta$ is the effective angle.
        Using the latter, we set the line of sewing $z(\rho)$ in (32$^*$), (33$^*$)
        at $z> 0$ in the form $\rho =
       R_{d} + (z-h_{d}/2)tg\theta(z)$. The condition $\theta_{0}=\theta(z=h_{d}/2)$ determines this angle at the beginning of the line of sewing.
             In calculations, we used the properties
      $ J_{l-1} (Q_{1}R_{d})=c_{l}J_{l+1} (Q_{1}R_{d})$ and $n_{l-1}(Q^{h}_{1}R_{d})=b_{l}n_{l+1} (Q^{h}_{1}R_{d})$, as well as the
       asymptotics of the Bessel and Neumann functions. Since $k_{z}+\kappa_{z}+Q_{1}tg\theta_{0} \ll k_{\rho}$,
       the relation $|I_{nuc}^{(1)}| \ll |I_{nuc}^{(2)}|$ is valid.

         Consider the case where $k_{z} \rightarrow \infty$.  Now, $k_{\rho}$ is small.
          The smallest $k_{\rho}$ are determined by the equality  $k^{(j)}_{\rho}=\mu_{l}^{(j)}/R_{d}$ (see (46$^*$)), where $l=l_{rot}=66$.
       Using the formulas \cite{y}
         \begin{equation}
        \mu_{l}^{(1)} \approx l+1.856\,l^{1/3} + 1.033\,l^{-1/3},
            \label{I20} \end{equation}
         \begin{equation}
        \mu_{l}^{(2)} \approx l+3.245\,l^{1/3} + 3.158\,l^{-1/3},
            \label{e18} \end{equation}
             \begin{equation}
        \mu_{l}^{(3)} \approx l+4.382\,l^{1/3} + 5.76\,l^{-1/3},
            \label{e19} \end{equation}
         we determine three first zeros of the Bessel function:  $\mu_{66}^{(1)}\approx 73.756$,
       $\mu_{66}^{(2)}\approx 79.895$, and $\mu_{66}^{(3)}\approx 85.134$.
            In order to calculate $I^{(1)}_{nuc}$ and $I^{(2)}_{nuc},$ we take into account that, at small $k_{\rho},$ the component
             $k_{z}=\sqrt{k^{2}-k^{2}_{\rho}}\approx k-k^{2}_{\rho}/2k \approx k$. Then we have
             \begin{equation}
           I_{nuc}^{(1)}(k_{z} \rightarrow \infty)
           \approx -f_{nuc}^{(1)}\frac{\sin{(kh_{d}/2)}}{4k},
                             \label{n1} \end{equation}
           \begin{eqnarray}
           f_{nuc}^{(1)} &=&  \int \rho d\rho
           \left [J_{l-1}(Q_{1}\rho) J_{l-1}(k_{\rho}\rho)\right. + \nonumber \\
           &+& \left. c_{l}J_{l+1}(Q_{1}\rho) J_{l+1}(k_{\rho}\rho)\right ].
     \label{n2} \end{eqnarray}
                     For three smallest $k^{(j)}_{\rho}$ ($j=1; 2; 3$), we determined numerically the following values:
            $f^{(1)}_{nuc}/R_{d}^{2}=0.0023$, -0.00049, and 0.00029.  By calculating integral (\ref{a33}), we have
              \begin{eqnarray}
        & &I_{nuc}^{(2)}(k_{z} \rightarrow  \infty) \approx \frac{\cos{\left [(\pi/\tilde{h}_{d}-k_{z})h_{d}/2\right ]
           J_{l-1}(Q_{1}R_{d})}}{(\pi/\tilde{h}_{d}-k_{z})^{2}ctg\theta_{0}} \nonumber \\
        &  & \times  R_{d}\left [J_{l-1}(k_{\rho}R_{d})+J_{l+1}(k_{\rho}R_{d}) \right ] + (k_{z}\rightarrow -k_{z}),
                             \label{n3} \end{eqnarray}
      which implies that $I_{nuc}^{(2)}$ is negligible:  $|I_{nuc}^{(2)}(k_{z} \rightarrow \infty)| \sim |I_{nuc}^{(1)}(k_{z} \rightarrow \infty)|/kR_{d}
           \ll |I_{nuc}^{(1)}(k_{z} \rightarrow \infty)|$.

 \subsection{Calculation of $F_{fi}^{el}$} \label{2.2}
        The electron part, $F_{fi}^{el}$, of the total matrix
         element (\ref{a19}) can be calculated analogously.
 The nonzero value of $F_{fi}^{el}$ is determined by the difference
   \begin{eqnarray}
       && \textbf{A}(\textbf{R} + \textbf{r}) -  \textbf{A}(\textbf{R}) =
        \textbf{e}_{\varphi}(\textbf{R})\cdot \left (\frac{\partial A_{\varphi}}{\partial \textbf{R}}\textbf{r} \right )
        + \nonumber \\ &+& \textbf{e}_{\rho}(\textbf{R})\cdot \left
        (\frac{\partial A_{\rho}}{\partial \textbf{R}}\textbf{r} \right )
        -  \textbf{A}(\textbf{R})\frac{y\sin{\varphi}+x\cos{\varphi}}{\rho} + \nonumber \\
      &+&  A_{\varphi}\frac{-y\textbf{i}+x\textbf{j}}{\rho}
        + A_{\rho}\frac{x\textbf{i}+y\textbf{j}}{\rho} + O(r^{2}),
     \label{e1} \end{eqnarray}
     where $\textbf{R}=(\rho\cos{\varphi},\rho\sin{\varphi},Z)$, $\textbf{r}=(x,y,z),$ and we took into account that, as $R$
     is much more than the atom size ($R \gg r$),
       \begin{equation}
      \textbf{e}_{\varphi}(\textbf{R}+\textbf{r})  \approx  \textbf{e}_{\varphi}(\textbf{R})
     \left (1- \frac{y\sin{\varphi}+x\cos{\varphi}}{\rho}\right ) - \frac{y}{\rho}\textbf{i} +\frac{x}{\rho}\textbf{j},
     \label{e2} \end{equation}
     \begin{equation}
      \textbf{e}_{\rho}(\textbf{R}+\textbf{r})  \approx  \textbf{e}_{\rho}(\textbf{R})
     \left (1- \frac{y\sin{\varphi}+x\cos{\varphi}}{\rho}\right ) + \frac{x}{\rho}\textbf{i} +\frac{y}{\rho}\textbf{j}.
     \label{e3} \end{equation}
  Representing $\textbf{A}$ in the form
   \begin{equation}
  \textbf{A} =  A_{\rho}(\rho,\varphi,Z)\textbf{e}_{\rho}+
 A_{\varphi}(\rho,\varphi,Z)\textbf{e}_{\varphi},
         \label{a43} \end{equation}
     we obtain
   \begin{equation}
      F_{fi}^{el} = \frac{i\hbar
      |e|n\tilde{c}}{m_{e}c\sqrt{N}}(I_{1}+I_{2}),
                \label{e4}  \end{equation}
       \begin{eqnarray}
           I_{1} &=&  \int d\textbf{R} e^{-il_{c}\varphi
    -ik_{z}Z}J_{l}(k_{\rho}\rho)\times \nonumber \\
    &\times & \left [A_{\rho} +  \partial A_{\varphi}/\partial\varphi + \rho\partial A_{\rho}/\partial\rho \right ]/\rho,
     \label{e5} \end{eqnarray}
     \begin{eqnarray}
           I_{2} &=&  n\int d\textbf{R}_{1}d\textbf{R}_{2} g(\textbf{R}_{1} - \textbf{R}_{2})
       e^{-il_{c}\varphi_{2}-ik_{z}Z_{2}} \times \label{e6}\\
     &\times & \frac{J_{l}(k_{\rho}\rho_{2})}{\rho_{1}}\left [A_{\rho}(\textbf{R}_{1}) + \frac{\partial A_{\varphi}(\textbf{R}_{1})}{\partial\varphi_{1}} +
     \frac{\rho_{1}\partial A_{\rho}(\textbf{R}_{1})}{\partial\rho_{1}}\right  ].
     \nonumber  \end{eqnarray}
           Using relations (72$^*$), (\ref{I64})--(\ref{I57}), we can verify that
      \begin{equation}
           I_{2} \approx I_{1}[S(k)-1]\frac{B(k_{\rho}R_{\infty},l)}{1-\tilde{R}_{d}/R_{\infty}},
     \label{e7} \end{equation}
             \begin{equation}
           I_{1} =  2\pi A_{m}\delta_{l,l_{c}}I_{el},
     \label{e8} \end{equation}
        \begin{eqnarray}
         & I_{el}&= \int dZ d\rho J_{l}(k_{\rho}\rho)e^{-ik_{z}Z}\left \{(1-l)a_{1}(\rho,Z)\right. -  \label{e9}  \\
           &-& \left. (1+l)a_{2}(\rho,Z) + \rho \partial /\partial\rho \cdot(a_{1}(\rho,Z) -
            a_{2}(\rho,Z)) \right  \},
     \nonumber  \end{eqnarray}
     where the integration is carried on over the volume occupied by helium. Analogously to (\ref{a33-2}), we divide $I_{el}$ into the sum of
      integrals over regions I and II. Then we have
       \begin{equation}
            I_{el} = I^{(1)}_{el} + I^{(2)}_{el},
                \label{e12} \end{equation}
         \begin{equation}
          I^{(1)}_{el}= \left (\int\limits_{-\infty}^{-h_{d}/2} + \int\limits^{\infty}_{h_{d}/2}\right )\frac{dZ}{8}
          e^{-ik_{z}Z-\kappa_{z}(|Z|-h_{d}/2)}I^{(1)}_{el,\rho}(Z),
                \label{e13} \end{equation}
       \begin{eqnarray}
          & I^{(1)}_{el,\rho}(Z)&= \int\limits_{0}^{\breve{R}_{d}} d\rho J_{l}(k_{\rho}\rho)
             \left [(1-l)J_{l-1}(Q_{1}\rho) - \right. \nonumber \\
           & - &  (1+l)c_{l}J_{l+1}(Q_{1}\rho)
           +Q_{1}\rho  J^{\prime}_{l-1}(Q_{1}\rho) -  \label{e14}  \\
           &-& \left. c_{l}Q_{1}\rho J^{\prime}_{l+1}(Q_{1}\rho)\right ],
             \nonumber   \end{eqnarray}
                       \begin{eqnarray}
           I_{el}^{(2)} &=& \int\limits_{R_{d}}^{\infty} d\rho I_{z}(k_{z},\rho)J_{l}(k_{\rho}\rho)
           \left [(1-l)n_{l-1}(Q_{1}^{h}\rho) -  \right.\nonumber \\
           &-& (1+l)b_{l}n_{l+1}(Q_{1}^{h}\rho) + \label{e15} \\
           &+& \left. Q_{1}^{h}\rho n^{\prime}_{l-1}(Q_{1}^{h}\rho) -
           b_{l}Q_{1}^{h}\rho n^{\prime}_{l+1}(Q_{1}^{h}\rho)\right ],
     \nonumber \end{eqnarray}
        where  $I_{z}(k_{z},\rho)$ is defined above, $\breve{R}_{d}=R_{d}+(|Z|-h_{d}/2)tg\theta(Z)$,
        $J^{\prime}_{l}(x)=\partial J_{l}(x)/\partial x$, and $n^{\prime}_{l}(x)=\partial n_{l}(x)/\partial x$.
     The integrals $I_{el}^{(1)}$ and $I_{el}^{(2)}$ are calculated analogously to $I_{nuc}^{(1)}$ and  $I_{nuc}^{(2)}$.
      Eventually, we get
       \begin{eqnarray}
      && I_{el}^{(1)}(k_{\rho} \rightarrow \infty)  \approx  -\frac{l J_{l-1}(Q_{1}R_{d})}{2k^{2}_{\rho}tg\theta_{0}}
       \sqrt{\frac{2}{\pi k_{\rho}R_{d}}} \times \nonumber \\
       &\times & \cos{\alpha_{l}(k_{\rho}R_{d})}\cos{(k_{z}h_{d}/2)}
       \left \{ 1- \frac{Q_{1}R_{d}}{2l J_{l-1}(Q_{1}R_{d})}\times\right. \nonumber \\
       \nonumber \\ &\times & \left.
       \left [ J^{\prime}_{l-1}(Q_{1}R_{d}) -
           c_{l}Q_{1}\rho J^{\prime}_{l+1}(Q_{1}R_{d}) \right ]\right \},
      \label{e16}  \end{eqnarray}
             \begin{eqnarray}
      && I_{el}^{(2)}(k_{\rho} \rightarrow \infty) \approx \frac{2l-2}{k_{\rho}}\sqrt{\frac{2}{\pi k_{\rho}R_{d}}}
           \sin{\alpha_{l}(k_{\rho}R_{d})}\times \nonumber \\
        && \times I_{z}(k_{z},R_{d}) = -k_{\rho}I_{nuc}^{(2)}(k_{\rho} \rightarrow \infty).
     \label{e17} \end{eqnarray}
     At $k \gg 1/h_{d},$ the integral $I_{el}^{(1)}$ is negligible: $|I_{el}^{(1)}|\sim |I_{el}^{(2)}|/k_{\rho}h_{d} \ll |I_{el}^{(2)}|$.
     Formulas (\ref{a34}) and (\ref{e16}) do not involve the condition $\Psi_{c}=0$ for the disk surface.
      For its consideration, we need to make replacements $\cos{(k_{z}h_{d}/2)} \rightarrow \sin{(k_{z}h_{d}/2)}$, $\cos{\alpha_{l}(k_{\rho}R_{d})}\rightarrow
      \sin{\alpha_{l}(k_{\rho}R_{d})}/k_{\rho}R_{d} \approx 1/k_{\rho}R_{d} \ll 1$ in the formulas, which decreases $I_{nuc}^{(1)}$ and $I_{el}^{(1)}$ still further.

   At small $k_{\rho},$ we obtain
        \begin{equation}
           I_{el}^{(1)}(k_{z} \rightarrow \infty)
           \approx -f_{el}^{(1)}\frac{\sin{(kh_{d}/2)}}{4k},
                             \label{n4} \end{equation}
           \begin{eqnarray}
           f_{el}^{(1)} &=&  \int\limits_{0}^{R_{d}} d\rho J_{l}(k_{\rho}\rho)
             \left [(1-l)J_{l-1}(Q_{1}\rho)\right. - \nonumber \\
             &-&  (1+l)c_{l}J_{l+1}(Q_{1}\rho)       +Q_{1}\rho  J^{\prime}_{l-1}(Q_{1}\rho) -\label{n5} \\
          &-& \left. c_{l}Q_{1}\rho J^{\prime}_{l+1}(Q_{1}\rho)\right ].
      \nonumber \end{eqnarray}
                            For $k^{(j)}_{\rho}$ with $j=1; 2; 3,$ we get numerically:
            $f^{(1)}_{el}/R_{d}=-0.17, 0.039$, and $-0.025$.  In addition,
              \begin{eqnarray}
         &&  I_{el}^{(2)}(k_{z} \rightarrow \infty) \approx \frac{\cos{\left [(\pi/\tilde{h}_{d}-k_{z})h_{d}/2\right ]
           }}{        (\pi/\tilde{h}_{d}-k_{z})^{2}ctg\theta_{0}} \times \nonumber \\
         &&  \times J_{l-1}(Q_{1}R_{d})J_{l}(k_{\rho}R_{d}) \times \label{n6} \\
           &&\times\left \{(1-l)n_{l-1}(Q_{1}^{h}R_{d}) - (1+l)b_{l}n_{l+1}(Q_{1}^{h}R_{d}) + \right. \nonumber \\
          && \left.  +Q_{1}^{h}R_{d} [J^{\prime}_{l-1}(Q_{1}^{h}R_{d}) -
           b_{l} J^{\prime}_{l+1}(Q_{1}^{h}R_{d})]\right \}+ \nonumber \\
           &+& (k_{z}\rightarrow -k_{z}) \sim \frac{I_{el}^{(1)}(k_{z} \rightarrow \infty)}{kR_{d}} \ll  I_{el}^{(1)}(k_{z} \rightarrow \infty).
                          \nonumber    \end{eqnarray}

     With regard for (\ref{a32a}), (\ref{a32b}), (\ref{e4})--(\ref{e8}), the performed calculations allow us to write finally
        \begin{equation}
      F_{fi}^{el} = \delta_{l,l_{c}}b(k_{\rho},l)\xi\sqrt{k_{\rho}}I_{el},
                \label{e20}  \end{equation}
    \begin{equation}
      b(k_{\rho},l) = \frac{m_{4}}{m_{e}}\left (\frac{1-\tilde{R}_{d}/R_{\infty}}{[S(k)-1]B(k_{\rho}R_{\infty},l)}+ 1
       \right ).
                \label{e21}  \end{equation}
                \subsection{Total matrix transition element $F_{fi}$} \label{2.3}
      We have
       \begin{equation}
      F_{fi} = F_{fi}^{el} + F_{fi}^{nuc} \approx \delta_{l,l_{c}}\xi\sqrt{k_{\rho}}
      [k_{\rho}I_{nuc} + b(k_{\rho},l)I_{el}],
                \label{e22}  \end{equation}
     whence the total probability of the creation of a c-phonon with
         ``momentum'' $k$ is
             \begin{eqnarray}
      & w_{fi} &= \sum\limits_{l_{c},n_{z},n_{\rho}}\delta w_{fi} = \nonumber \\
      &=&  \frac{2\pi}{\hbar}\sum\limits_{l_{c},n_{z},n_{\rho}}|F_{fi}|^{2}\delta (E_{f}-E_{i}^{(0)}-\hbar\omega) =
       \label{e27}  \\
       &=& \frac{2\pi}{\hbar}\sum\limits_{n_{z},n_{\rho}}|\xi|^{2}k_{\rho}[k_{\rho}I_{nuc} + b(k_{\rho},l)I_{el}]^{2}\delta (E_{c}(k)-\hbar\omega).
      \nonumber \end{eqnarray}
      Moreover, $k^{2}=k_{\rho}^{2}+k_{z}^{2}$ in all formulas. The main contribution to sum (\ref{e27})
       is given by the regions ($k_{z}\rightarrow 0$, $k_{\rho}\rightarrow k$) and
       ($k_{\rho}\rightarrow 0$, $k_{z}\rightarrow k$). We now determine these contributions.

       a) Region of small $k_{z}$ (here, $k_{\rho}\rightarrow k$). According to
       (\ref{a33-2})--(\ref{a36}) and (\ref{e12}), (\ref{e16}), (\ref{e17}), we have
        $I_{nuc}(k_{\rho}\rightarrow \infty)\approx I_{nuc}^{(2)}(k_{\rho}\rightarrow \infty)$,
         $I_{el}(k_{\rho}\rightarrow \infty)\approx I_{el}^{(2)}(k_{\rho}\rightarrow \infty)$.
       In (\ref{e27}), we now pass from the sum $\sum\limits_{n_{\rho}}$ to the integral
       $\sum\limits_{n_{\rho}} \rightarrow \frac{R_{\infty}-R_{d}}{\pi}\int dk_{\rho}$ and then, with the help of the
       relation $kdk=k_{\rho}dk_{\rho},$ to an integral over $k$. Due to the $\delta$-function, this
       integral is easily calculated. After some transformations, we obtain
        \begin{equation}
       w_{fi}(k_{z}\rightarrow 0) = S_{1}\frac{2b_{0}^{2}|\xi_{0}|^{2}h_{d}^{2}(R_{\infty}-R_{d})}{R_{d}\hbar k^{2}\partial E/\partial
       k},
       \label{e29}  \end{equation}
        \begin{eqnarray}
      &&S_{1}(k)=\frac{8(l-1)^{2}}{\pi h_{d}^{2}}\sum\limits_{n_{z}}
       \frac{k^3}{k^{3}_{\rho}}\sin^{2}{\alpha_{l}(k_{\rho}R_{d})}I_{z}^{2}(k_{z},R_{d})\approx
               \nonumber \\ & \approx &
       (2/\pi)\left [(l-1)J_{l-1}(Q_{1}R_{d})\sin{\alpha_{l}(k R_{d})}\right ]^{2}\times
        \nonumber \\ &\times  &  \sum\limits_{n_{z}} \left
       \{\frac{\sin{(\frac{\pi h_{d}}{2\tilde{h}_{d}}-n_{z}\frac{\pi h_{d}}{H-h_{d}})}}
       {\frac{\pi h_{d}}{2\tilde{h}_{d}}-n_{z}\frac{\pi h_{d}}{H-h_{d}}} + (k_{z} \rightarrow -k_{z})\right \}^{2}\approx
         \nonumber \\ & \approx & \frac{4.3 (H-h_{d})}{\pi h_{d}}
         \left [(l-1)J_{l-1}(Q_{1}R_{d})\sin{\alpha_{l}(k R_{d})}\right ]^{2},
     \label{e30} \end{eqnarray}
           where  $n_{z}=\pm 1,\pm 2,\ldots$, $b_{0}= b(B=1,\tilde{R}_{d}=R_{d})=\frac{m_{4}}{m_{e}}\frac{S(k)-R_{d}/R_{\infty}}{S(k)-1}$,
           $\xi_{0}= \xi(B=1,\tilde{R}_{d}=R_{d})$, and the value of $k$ is determined from the condition $E_{c}(k)\equiv E(k)=\hbar\omega.$
            For $k_{z},$ we used the quantization laws
            (\ref{I42a}),  which assumes the zero boundary conditions on the container walls.

     b) Region of small $k_{\rho}$. With the use of $k_{z}=\frac{2\pi n_{z}}{H-h_{d}}$ (see (\ref{I42a})) and $k_{\rho}^{(j)}=\mu_{l}^{(j)}/R_{d}$, relations
         (\ref{a33-2}), (\ref{n1})--(\ref{n3}), (\ref{e12}), (\ref{n4})--(\ref{n6}) yield
      \begin{eqnarray}
      && \sum\limits_{n_{z}}f(k)\delta (E(k)-\hbar\omega) =
       \frac{H-h_{d}}{2\pi}\int\limits_{-\infty}^{\infty}dk_{z}f(k)\times \nonumber \\
       && \times\delta   (E(k)-\hbar\omega) = \frac{H-h_{d}}{\pi}\frac{k}{k_{z}}\frac{f(k)}{\partial E/\partial k},
            \label{e32} \end{eqnarray}
       \begin{equation}
       w_{fi}(k_{\rho}\rightarrow 0) = S_{2}\frac{2(H-h_{d})R_{d}b_{0}^{2}|\xi_{0}|^{2}}{\hbar
       k^{2}\partial E/\partial k},
       \label{e33}  \end{equation}
        \begin{eqnarray}
      && S_{2}(k)=\sum\limits_{j= 1,\ldots} \frac{k}{k_{z}}\frac{k_{\rho}^{(j)}}{R_{d}}\frac{1-f}{1-f_{j}}\times \nonumber \\
      &&\times\left |\frac{S(k)-f}{S(k)-1+B^{-1}(k_{\rho}^{(j)}R_{\infty},l)(1-f_{j})}
      \right | \times    \nonumber\\
      && \times\left \{ \frac{k_{\rho}^{(j)}kI_{nuc}(k_{\rho}^{(j)})}{b_{0}} + kI_{el}(k_{\rho}^{(j)}) \times \right. \label{e34} \\
      && \left. \times \frac{S(k)-1+(1-f_{j})B^{-1}(k_{\rho}^{(j)}R_{\infty},l)}{S(k)-f}
       \right \}^2\equiv  \nonumber \\ && \equiv  a(k,l)\sin^{2}{\left (\frac{kh_{d}}{2}\right )}. \nonumber
     \end{eqnarray}
      where $f=R_{d}/R_{\infty}$, $f_{j}=\tilde{R}_{d}(k_{\rho}^{(j)})/R_{\infty}$.

           Formulas (\ref{e30}) and (\ref{e34}) include $\sin{\alpha_{l}(k R_{d})}$ and $\sin{(kh_{d}/2)}$.
          Due to the zero boundary conditions, $\sin{\alpha_{l}(k_{\rho}R_{d})}\approx \pm 1$, whence $\sin{\alpha_{l}(k R_{d})}\approx \pm 1$.
          In view of the same conditions, we set $\sin{(kh_{d}/2)}\approx \pm 1$.

             Using (\ref{a32a}), (\ref{a32b}), (75$^*$), and the formula for the volume of helium $V=\pi R_{\infty}^{2}H$,
       we obtain finally:
        \begin{eqnarray}
        & &    w_{fi}\approx   w_{fi}(k_{z}\rightarrow 0) +  w_{fi}(k_{\rho}\rightarrow 0) \approx  \nonumber \\
       & &\approx \frac{2\pi^{2}(1-h_{d}/H)n\hbar (eA_{m})^{2}|S(k)-f|}{c^{2}k^{2}m_{e}^{2}\partial E/\partial k}\times
         \label{e37}  \\ &&
        \times\left (\frac{8.6h_{d}}{\pi R_{d}}\left [(l-1)J_{l-1}(Q_{1}R_{d})\right ]^{2}
        +\frac{2f\cdot a(k,l)}{1- f}\right ).
         \nonumber \end{eqnarray}

        We note that the main contribution to $w_{fi}(k_{\rho}\rightarrow 0)$ and $w_{fi}(k_{z}\rightarrow
        0)$ is given by the electron part of $F_{fi}$, and
        the transition probability $w_{fi}$ is determined by the quantities $w_{fi}(k_{z}\rightarrow 0)$ and
        $w_{fi}(k_{\rho}\rightarrow 0),$ i.e., by the creation of c-phonons with the smallest $k_{z}$ and large
        $k_{\rho}$ and with the smallest $k_{\rho}$ and large
        $k_{z}$ (almost plane c-phonons). We now evaluate these contributions quantitatively for the roton line
        and conditions of the experiment \cite{svh1,svh3}. According to \cite{I}, the relation $J_{l-1}(Q_{1}R_{d}) \approx 1/27.831$ is valid at $l=l_{rot}=66$.
        To calculate $a(k,l)$ in (\ref{e34}), we use the roton value $S(k)\approx 1.3$ (for $T \lsim 1.4\,K$). Moreover, we consider that
         the relation $k_{z}/k \approx 1$ holds as $k_{\rho}\rightarrow 0,$
   $\tilde{R}_{d}\approx 1.5R_{d}$ at $j=1$ \cite{I}, and the quantity $\tilde{R}_{d}$ decreases down to $R_{d}$ with increase in $j$.
     We calculated values of $B(k_{\rho}R_{\infty},l=66)$ numerically by formula (\ref{I57}): $B(k^{(1)}_{\rho}R_{\infty},66)\approx 0.91$,
            $B(k^{(2)}_{\rho}R_{\infty},66)\approx 0.93$, and $B(k^{(3)}_{\rho}R_{\infty},66)\approx 0.94$.
          The contribution of the following $k_{\rho}$ (with $j>3$) to $a(k,l)$ is small. Whence we find $a(k_{rot},l)\approx 0.2$.
          Moreover, since $B\approx 1,$ the functions $a(k,l)$ and
          $S_{2}$  are almost independent of $k$.
          But if $k$ are such that $S(k)\rightarrow f$,
            then  the coefficient $a(k,l)$ increases by several orders of magnitude, approximately as
           $|S(k)-f|^{-1}$, and becomes $\gg 0.2$. But, for the probability $w_{fi}$, such a growth  is cancelled by the factor
         $|S(k)-f|$ in (\ref{e37}) in front of the large parentheses.

         Using these numbers, the relation $R_{\infty}\approx H/2 \approx 2.1\,cm$, and (\ref{e34}), we obtain that the probability of the creation of
         c-phonons with large $k_{\rho}$ is $\sim 5$ times more than that for the almost planes ones (with small $k_{\rho}$).
        Thus, a resonator creates c-phonons mainly of two ``extreme'' types: almost completely circular and almost plane.
        The high probability of the creation of c-phonons with an almost plane structure is a somewhat unexpected result.
        At the same time, it is clear that the completely plane phonons cannot be created due to the
        angular momentum conservation law: a phonon must possess a certain ``twist'' ($l\neq 0$, $k_{\rho}\neq 0$)
        in order to carry away the angular momentum of a c-photon.

      \section{Width of the roton  absorption line}
       As was mentioned above, the spectrum of the SHF emission of a disk resonator contains the
       very narrow absorption line at the roton frequency ($\hbar \omega  = \Delta_{rot} =8.65\,K\cdot k_{B}$).
       Let the line width be the distance between the points, which located on both sides from the line center,
       and for which the intensity of a signal is about 0.8 of the background one (the signal far from the line). Then it follows from
       the experiment  \cite{svh1,svh3} that the line width is about $50\,\mbox{kHz}$ at $T=1.8\,K$ and decreases, at lower
       $T,$ to the resolving power of a spectroscope
       ($\simeq 30\,\mbox{kHz}$). Moreover, the width stops to decrease at $T \lsim 1.6\,K$ and approaches a constant $\simeq 30\,\mbox{kHz}$.
        The minimum experimental temperature $T=1.4\,K$, but formula (\ref{e37}) is applicable, if there are no c-rotons in the initial state of
        helium, i.e., at $T=0$. Because the width does not depend on $T$ already at $T \lsim 1.6\,K$,
          we may assume that the width at $T=0$ is the same as that at $T=1.4\,K$.

        The creation of a c-phonon  (or a
        c-roton) is caused by the c-photon
        $\rightarrow$ c-phonon transition. In order to calculate its probability,
        we should divide $w_{fi}$ (\ref{e37}) by the number of c-photons in a resonator, $N_{phot}.$ We can calculate the latter
         by dividing the total energy of the EM field of the resonator,
         \begin{equation}
       W =  \int dV
       \frac{\textbf{D}\textbf{E}+\textbf{B}\textbf{H}}{8\pi} = \int dV
       \frac{\varepsilon_{\perp} \dot{\textbf{A}}^{2}/c^{2}+(rot \textbf{A})^{2}}{8\pi},
     \label{a53} \end{equation}
         by the energy of a c-photon, $\hbar\omega$. Since the field outside of the disk
         is weak, we will consider only the field inside the disk determined by formulas
         (17$^*$) and (18$^*$). Since
         \begin{equation}
           \int\limits_{0}^{R_{d}}\rho d\rho J_{l-1}^{2}(Q_{1}\rho)  \approx
           \frac{R_{d}^2}{2}\left [J_{l-1}^{\prime}(\mu_{l-1}^{(1)}) \right ]^2 \approx 2\cdot 10^{-3}R_{d}^{2}
       \label{a54} \end{equation}
         and  $\cos{(\pi h_{d}/2\tilde{h}_{d})}=1/8$, we can obtain
      \begin{eqnarray}
       W &\approx & 10^{-3}A_{m}^{2}R_{d}^{2}h_{d}\left [
       \left (\frac{\varepsilon_{\bot} \omega^2}{c^2}+\frac{Q_{1}^{2}}{2}\right )\left (1+\frac{\tilde{h}_{d}}{4\pi h_{d}}\right )
       \right. + \nonumber \\
        &+& \left. \frac{\pi^{2}}{\tilde{h}_{d}^{2}}\left (1-\frac{\tilde{h}_{d}}{4\pi h_{d}}\right )\right ],
     \label{a55} \end{eqnarray}
    where the small terms contributing to $W$ are omitted. Using (\ref{a55}), it is easy to evaluate
       the total energy and the number of c-photons in the pumping band
     (${ \sss\triangle} \nu_{pump} \simeq 50\,\mbox{kHz}$) for the roton mode ($l=66$) and the roton frequency:
          \begin{equation}
       W \approx 9.2A_{m}^{2}h_{d}(R_{d}/0.95cm)^{2}  \approx 3.87\cdot 10^{8}\,\mbox{eV},
     \label{a55-2} \end{equation}
              \begin{equation}
       N_{phot} = \frac{W}{\hbar\omega} \approx 5.19\cdot 10^{11}.
     \label{a56} \end{equation}

      In a vicinity of the roton minimum, we have $\partial E(k)/\partial k =\hbar^2 |k-k_{rot}|/m_{rot}$
     with   $m_{rot}\approx 0.165m_{4}$,  $k_{rot}=1.93\,\mbox{\AA}^{-1}$.  Then the probability $ \tilde{w}_{fi} = w_{fi}/N_{phot}$
     of the c-photon $\rightarrow$ c-roton process for  the experimental width ${ \sss\triangle} \nu =
        30\,\mbox{kHz}$ is
      \begin{eqnarray}
      \tilde{w}_{fi} & \approx &
      \frac{4\pi^{2}}{9.2} \frac{(1-h_{d}/H)ne^{2}\omega m_{rot}|S(k)-f|}{m_{e}^{2}k^{2}c^{2}h_{d}|k-k_{rot}|}
         \times    \nonumber  \\  & \times &
       \left (\frac{4.3h_{d}}{\pi R_{d}}\left [(l-1)J_{l-1}(Q_{1}R_{d})\right ]^{2}
        +\frac{fa(k,l)}{1- f}\right ) \times \nonumber \\
        &\times &   \left (\frac{0.95 cm}{R_{d}}\right )^{2}    \approx    3.36\cdot 10^{-7}\omega_{rot}.
          \label{a57} \end{eqnarray}
           This process weakens the flow of photons propagating from the resonator. We now determine the absorption line width.
           To this end, we consider properties of the resonator. In the experiment \cite{rub3}, the pumping signal
           with the frequency band ${ \sss\triangle} \nu_{pump}$ and the power
           $w^{0}_{pump}\simeq 10^{-3}$W was firstly switched-on. But, as a result of losses, the resonator received $w_{pump}\simeq 10^{-4}\div 10^{-5}$W.
           The resonator accumulates and amplifies the pumping signal,
           but, in this case, the energy losses in the resonator increase also, until a stationary equilibrium state is established.
           In this state, the losses in the resonator on the emission are equal to the pumping.
         In other words, the condition of equilibrium for the EM field of the resonator
          with frequencies in the pumping band is the equality of the pumping energy flow and the losses:
            \begin{equation}
      w_{pump} =N_{phot}\hbar \omega/\tau_{ren},
      \label{eq1} \end{equation}
         where  $N_{phot}$ and $\tau_{ren}$ are, respectively, the
        number of c-photons in the resonator and the mean emission time of a c-photon by the resonator (obviously, it is also the period of renewal
        of EM modes of the resonator).
        By the value of $w_{pump},$ we can estimate the electrical signal formed by  an antenna catching photons emitted by the resonator
        as $w\simeq 4\cdot 10^{-8}$W. It follows from the distribution of the stationary EM field of the resonator \cite{I} that
        this field induces in the antenna the signal $w\lsim 4\cdot 10^{-21}$W,  weaker by 13 orders.
        In other words, the signal generated by
        the antenna is due to photons emitted by the resonator, rather than to the stationary EM field of the latter.

         When the frequency  of the EM field approaches the roton one, then, according to (\ref{a57}),
        the probability of the emission of c-rotons by the resonator becomes large. We now consider the frequency interval
        ${ \sss\triangle} \nu_{0} = { \sss\triangle} \nu_{pump}/100 \simeq 0.5\,\mbox{kHz}$,
        which is much less than the line width and the pumping band but contains a macroscopic number of quanta.
        If the emission of c-rotons occurs, then the condition of equilibrium for the band ${ \sss\triangle} \nu_{0}$ takes the form
        \begin{equation}
      0.01w_{pump} = 0.01N_{phot}\hbar \omega/\tau_{ren} + N^{0}_{rot}\triangle_{rot}/\tau_{em},
      \label{eq2} \end{equation}
      where  $\tau_{em}=1/\tilde{w}_{fi}$ is a duration of the emission of a c-roton by c-photon of the resonator, and $N^0_{rot}$ is the
      number of c-rotons emitted for a time interval $\tau_{em}$ by c-photons from the band ${ \sss\triangle} \nu_{0}$.
      In this case, the losses of the resonator are separated into the channels of emission of c-photons and c-rotons. Respectively, the flow of emitted
      c-photons decreases, which is manifested in the resonator spectrum as the absorption. As was mentioned above,
      the signal is equal to $80 \%$ of the background one on the edge of lines. Hence, $20 \%$ of the losses of the resonator are transferred into c-rotons,
       whereas $80 \%$ pass into c-photons. Thus, each c-photon of the resonator in the frequency band ${ \sss\triangle} \nu_{0}$ for the time
       $\tau_{ren}$ emits a c-photon with a probability of 0.8 and a c-roton with a probability of 0.2. This yields
      $\tau_{em}=4\tau_{ren}$, which allows us to write the following condition for the line edge:
      \begin{equation}
      \tilde{w}_{fi} \equiv \frac{1}{\tau_{em}}= \frac{1}{4\tau_{ren}}.
      \label{lin} \end{equation}
           Relations (\ref{a56}), (\ref{eq1}) yield
         $\tau_{ren}\simeq 6.25 (10^{-7} \div 10^{-6})\,$s, which corresponds to the experimental value
          $\tau_{ren}\simeq 10^{-6}\,$s.  Using the last value of
          $\tau_{ren}$ and
          (\ref{lin}), we obtain $\tilde{w}_{fi}$ for the line edge: $\tilde{w}_{fi} = 1/4\tau_{ren} \simeq 2.5\cdot 10^{5}\,
          \mbox{s}^{-1}\approx 2.2\cdot 10^{-7}\omega_{rot}$, which is only by a factor of 1.5 less than the theoretical value of $\tilde{w}_{fi}$
         (\ref{a57}).

    \begin{figure}[h]
 % \vspace*{7.5cm}
\centerline{\includegraphics[width=85mm]{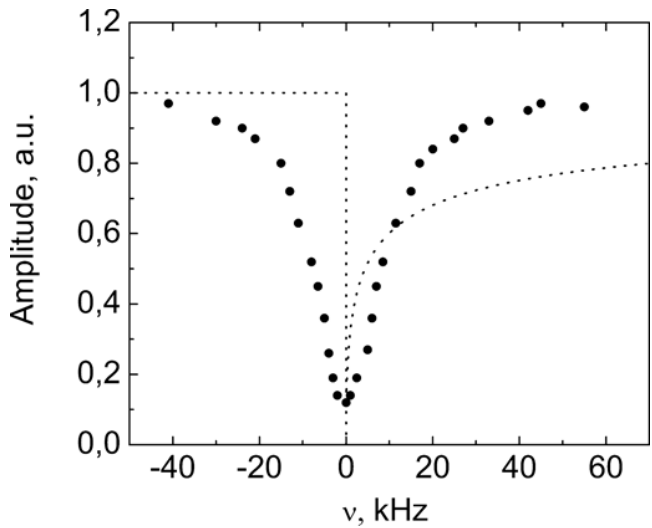}}
\caption{Intensity of the signal received by the antenna {\it vs}
the frequency near a narrow roton absorption line, $T=1.63\,$K.
$\bullet\bullet\bullet$
--- experiment \cite{svh1}, dotted line --- theory (by the present work).
The frequency is equal to that indicated in the figure plus
$175.7\,\mbox{GHz}$ (this number is the roton energy at
$T=1.63\,$K). }
\end{figure}
 %  \noindent
         The theoretical and experimental lines are shown in Fig.~1. At $\nu > \nu_{rot}$, the amplitude of the former is given by the formula
          $a\approx \left (1+\sqrt{\frac{4.4\,kHz}{\nu -\nu_{rot}}}\right )^{-1}$.
                     In this case, the error of the line width is about one order; it related to the approximate character of the solutions for the WF of a
      c-phonon  (not completely correct consideration of boundary conditions) and for the EM field of the resonator, as well as to
       the neglect of a deviation of a symmetry of the system far from the disk (antenna, container's wall, etc.)
       from the cylindrical symmetry.
              As is seen from Fig.~1, the theoretical line corresponds approximately to the experimental one by
           width, but  does not by shape. The density of states
       $\varrho (E) = \int \delta(E(k)-\hbar\omega)d\textbf{k} = \frac{4\pi k^{2}}{\partial E/\partial k}$
       is high in the region above  the energy $\Delta_{rot}$ of a roton
              ($\varrho (E) = \frac{4\pi k^{2}m_{rot}}{\hbar^{2}|k-k_{rot}|}\rightarrow \infty$
       at $k\rightarrow k_{rot}$).
        Below $\Delta_{rot},$ no roton states are present, and the density of states decreases
        sharply to a value corresponding to the linear phonon curve:
        $\varrho (E) = \frac{4\pi k^{2}}{u_{1}\hbar}$. Therefore, for the c-photon $\rightarrow$ c-roton transition,
        the line must sharply fall on the side of lower frequencies, whereas the experimental line is almost symmetric.
        This difference  means that the process is more complicated and involves at least two particles, rather than one particle
        (c-photon $\rightarrow$ c-roton),
        as was accepted above. In particular, one more c-phonon can be created, or
        a part of the angular momentum of a c-photon can be absorbed by the disk. We hope to clarify the mechanism
         in the future investigations.

                 It is necessary also to take into account that a spreading of the c-roton energy which is due to the interaction
          between quasiparticles can also contribute to the line width.
        However, this leads to a very large line width of the order of magnitude of the width $\sim 0.1\Delta_{rot}$ in neutron experiments.
         This value is larger by 6 orders than the width of the narrow SHF line and by 4 orders than the mode width.
         Therefore, this spreading cannot be observed for a single mode, but it is revealed as the ``pedestal'' (base)  on many modes \cite{svh3}.

        The number of photons approaching the receiver is diminished not only due to the creation of c-rotons by the field of the resonator,
        but also due to their creation by c-photons already emitted by the resonator. The time-of-flight of these photons to
        the receiver is $\sim 4\lambda/c_{h} \approx 3\cdot 10^{-11}\,\mbox{s}\simeq  10^{-5}\tau_{em},$ which is less by 5 orders than the duration of the emission
         of a c-roton by a photon of the resonator. In this case, the probabilities of the emission of a c-roton for a standing photon of the resonator and a running photon
          are of the same order, because a standing c-photon is simply a superposition of two radial c-photons propagating toward each other.
          Therefore, the line is formed not due to the absorption of running photons, but, as we have assumed,
           due to the absorption of ``standing'' photons of the resonator.

        \section{Discussion}

        While calculating formula (\ref{a57}), we used the solution for the EM field \cite{I}, for which the sewing in the corner
         region ($|z| \geq h_{d}/2, \rho =R_{d} \div \infty$) is approximate. In this region, the EM field is weak. But, nevertheless, it
         influences the value of $\tilde{w}_{fi}$. To evaluate this influence, we
         calculated the probability $\tilde{w}_{fi}$ for two other distributions of the EM field in the corner region:
         $\textbf{A}_h =0$ and  $\textbf{A}_h =   A_{m} e^{i(l\varphi - \omega t)} \left [a_{1}(\rho,z)(\textbf{e}_{\rho}+
  i\textbf{e}_{\varphi})+a_{2}(\rho,z)(-\textbf{e}_{\rho}+
  i\textbf{e}_{\varphi})\right ]$, $\{a_{1}(\rho,z),a_{2}(\rho,z)\} =
   \frac{1}{8}J_{l-1}(Q_{1}R_{d}) e^{-\kappa_{z}(|z|-h_{d}/2)} \times \\
   \times\{n_{l-1}(Q^{h}_{1}\rho),b_{l}n_{l+1} (Q^{h}_{1}\rho) \}$.
   The latter can be sewed with solutions in other regions, but it does not completely satisfy the equation. Moreover, if we use
   $Q_{1}$ instead of $Q^{h}_{1}$, then the equation is satisfied, but the sewing cannot be executed.  The value of $\tilde{w}_{fi}$ for such solutions differs from
    (\ref{a57}) by several times; therefore, $\tilde{w}_{fi}$ is not very sensitive to the sewing.
    The analysis indicates that it is of importance to correctly set the solution in the corner region near the joint
     with adjacent regions, and it is not so  important in bulk.
   Solution (32$^*$), (33$^*$) satisfies this condition, so that its use is justified.

        As was mentioned above, the probability  (\ref{a57}) of the c-photon $\rightarrow$ c-roton
      process grows strongly
         at $\partial E(k)/\partial k \rightarrow  0$, i.e., near the points of an extremum of the dispersion curve
      $E(k)$. This peculiarity explains why the narrow line is observed
      namely at the roton frequency and predicts the possibility to find one more line at the frequency of the maxon
      maximum, $\nu_{max} \approx
       287 \pm 2\,\mbox{GHz}$  \cite{and} (up to now, the
        frequencies $\nu \approx  40 \div
       200\,\mbox{GHz}$ were studied). Let us substitute the maxon parameters ($k_{max}=1.12\,\mbox{\AA}^{-1}$, $m_{max}\approx 0.54m_{4} $, $S(k)\approx 0.3$)
       in (\ref{a57}), and let us take into account that, in the denominator, the number $9.2\sim \omega^2$, and, according to (19$^*$) and (20$^*$), it should be $l \sim\omega$ for the resonance mode.
        We obtain $\tilde{w}_{fi} \approx 5.9\cdot 10^{-7}\omega_{max}$. Thus, without regard for an additional factor (see above),
        we have that
               if $\tau_{ren}$ is approximately identical for the maxon and roton lines,
          then the width of the maxon line must be larger by a factor of 1.7 (at $T=0$) than the width of the roton line.

    The peculiarity at $\partial E(k)/\partial k \rightarrow  0$ is well known in solid-state physics
    as the Van Hove singularity.  At $|\partial E (\textbf{k})/\partial
      \textbf{k}| \rightarrow 0,$ the states of c-rotons falling in the small given energy interval are strongly concentrated.
      Respectively, the transition probability in this energy interval sharply increases.
         In crystals, the narrow lines of light absorption \cite{kris1} and neutron scattering \cite{kris2} were registered a lot of times.
              However, the width of the latter is larger by several orders than those of SHF lines in helium and corresponds to the pedestal.
      Thus, the narrow SHF line in helium is related to the Van Hove singularity, like the lines of crystals, but its widening is caused by
      another mechanism.

       In neutron experiments with liquid helium, the analogous very
       narrow peaks must be observed on the scattering curve $S(k=const,\omega)$ at the frequencies of the roton and maxon extrema.
        However, the high error of neutron measurements ($\delta\omega \approx  0.1\,K$) does not allow one to register these
       peaks.

            In addition to the processes considered above, one more channel is possible: c-photon
       $\rightarrow$ p-phonon + the transfer of a momentum to the disk and the transfer of an angular momentum
       to the disk or for the creation of a vortex rotating around the disk.
       Such processes must be less probable, since the greater the number of quasiparticles participating in a process, the less is
       the probability of the process.
       Moreover, what is more important, the overlapping of the wave
       functions of the EM field and a p-phonon is slight due to different symmetries.

        In order to explain the appearance of the narrow line
        of absorption, the authors of work \cite{svh3} proposed to consider the following process: p-photon
       $\rightarrow$ p-roton + the transfer of a momentum to helium as a whole. In our opinion,
       the transfer of a momentum to the disk can be more probable in such an approach,
       since the disk is a tougher system as compared with
       helium. The approach with a plane photon and
       a plane phonon is the main alternative to the above-considered process with c-rotons.
        But the latter is, apparently, more probable by two reasons. \\
       1) As was mentioned above, the line is formed by the stationary EM field of the resonator, rather than by running photons.
       The EM field outside of the resonator can be represented as a
       superposition of plane waves (wave packets). But the total EM field of the resonator is a sum of c-photons
           localized outside and inside the disk (see \cite{I}). At the creation of a c-roton, a c-photon disappears \textit{as a whole} inside and outside of the disk.
         But, due to different values of $\varepsilon$
         of helium and the disk, such a c-photon cannot be presented in the form of a superposition of photons which are plane in both helium and the disk
         (a photon is not plane or in helium, or in the disk). In other words, we must be based in the input equations
         on the \textsl{circular} field of the resonator. In addition, the properties of this field are unlike those
         of plane waves, because the field of the resonator sharply drops with increase in the distance to the resonator. \\
        2) In addition, the approach with p-photons requires the expansion of the field in multipoles. Here, the main contribution is given by the
        term with the dipole moment (DM) $\textbf{d}_{r}$ of a roton.
        It was assumed in work \cite{svh3} that such a stationary DM
        arises in a roton due to the mutual polarization of atoms, and the probability of the p-photon $\rightarrow$
        p-phonon transition is proportional to $d_{r}$. It was also proposed in
        \cite{lt1} that a vortex ring possesses an intrinsic DM.
        However, both
        quasiparticles create the reciprocal motion of atoms: a part of atoms
        moves forward, but the similar part moves backward. In this case, the separated direction is set by the velocity of a quasiparticle. Therefore, the
        appearance of a DM is related to the asymmetry of a quasiparticle relative to the
        forward-backward directions. The question about the presence of such an asymmetry can be clarified with the help of the following
          reasoning (which belongs to Yu.\,V. Shtanov). By definition, the stationary DM of a quasiparticle
          is equal to
           \begin{eqnarray}
  & &\textbf{d}_{qp}=\int d{\bf R}_{1}d{\bf R}_1^{(1)} d{\bf R}_1^{(2)}\ldots
     d{\bf R}_{N}d{\bf R}_{N}^{(1)} d{\bf R}_{N}^{(2)}\Psi_{qp}^{*}\Psi_{qp}\cdot \nonumber \\
    &  \cdot & e(\textbf{R}_{1}^{(1)}+\textbf{R}_{1}^{(2)}-2 \textbf{R}_{1}\ldots
     +\textbf{R}_{N}^{(1)}+\textbf{R}_{N}^{(2)} -2\textbf{R}_{N}),
     \label{vr}     \end{eqnarray}
     where  $\Psi_{qp}=\Psi_{qp}({\bf R}_{1},{\bf R}_1^{(1)},{\bf R}_1^{(2)},\ldots
     {\bf R}_{N},{\bf R}_{N}^{(1)},{\bf R}_{N}^{(2)})$ is the WF of a quasiparticle. Let us make inversion of time
        $t\rightarrow -t$.  In this case, we have \cite{land3} $\Psi \rightarrow \Psi^*$,
     and DM (\ref{vr}) is not changed. But the DM must be directed along
     the velocity of a quasiparticle, i.e., it should change the sign.
     This implies that the DM is zero. It was considered in \cite{lt1} that, by the
     CPT-theorem, the charges also change their signs at $t\rightarrow -t$, which gives $\textbf{d}_{qp} \neq 0$.
       However, the change $t\rightarrow -t$ in the equation can be performed formally
         without any connection with the time arrow. Then
         the charges conserve their signs at $t\rightarrow -t,$ and the DM turns out zero.

         The real quasiparticle is a wave pocket, but it is clearly that in this case $\textbf{d}_{qp} = 0$ also.

      Such a consideration is not valid if the state of a quasiparticle is degenerate.
      In other words, at given $E$ and $\textbf{k},$ there are the states with DMs
      $\textbf{d}_{qp}$ and $-\textbf{d}_{qp}$ which can transit to each other at $t\rightarrow
      -t.$
       However, we have no reasons to consider that a roton or a ring has such a degeneration.
       For $\textbf{d}_{qp} \neq 0,$ there appears another possibility, if the reflected state is
       unstable and transits in a stable one with the inverse DM. This internal irreversibility can be related
       to the ordering of deformations of the electron shells
       of atoms induced by the interaction with neighbors. For clearness, we note that this is similar to a
       flag on a moving car. Such a flag points out always the direction opposite to one of the motion. At
       $t\rightarrow  -t,$ we obtain a flag indicating the direction of motion,
       i.e., we obtain the unstable state. A similar structure of a ring or
       a roton is possible in principle. But it is improbable, especially for a roton representing reciprocal oscillations of the density.
       It is of importance that such a change of the symmetry, like that at $t\rightarrow
      -t$, will happen at the reflection of a quasiparticle from the wall. At each reflection, the quasiparticle must loss energy.
       Hence, it will be unstable. Such an instability is possible for rings, and it could explain why the rings
       are not discovered in the spectrum of quasiparticles or by the contribution to the heat capacity till now.
       But such an instability seems impossible for phonons and rotons.
         Thus, a roton has no stationary DM. This is also the case for a ring, most probably.
         This reasoning implies that, while explaining the narrow line, the circular symmetry of the problem
         must be taken into account already in the input equations.

       \section{The line spectrum of the liquid $^{4}He$}
                   As is known from the general theorems of quantum
          mechanics, a many-particle
          system located in a finite volume possesses a discrete energy
spectrum.
           Liquid $^4$He in a vessel is the system of this kind.
           The energy levels of He~II can be determined from the WF of
         He~II calculated for the zero boundary conditions.
          From here, it is obvious that the real energy spectrum of He~II is not a
       Landau continuous curve,
       but it is a collection of separate disconnected points very densely lying on this curve. According to (\ref{I42a}), (\ref{I42b}), the observed
       roton line \cite{svh1,svh3} consists of $\sim 10^5$ individual roton lines. If the experiment will be executed with a film of helium
       $\sim 100\,\mbox{\AA}$ in thickness, the distances between lines increases by 6 orders of magnitude, and they will become resolvable.
       In this case, instead of a single roton
       line, we will measure many lines in the wide range of the frequencies: $\nu = 0\div 2\Delta_{rot}/2\pi\hbar$.
      %   , but not only at the roton one.
        Thus, it will be possible to observe for the first time the
       line spectrum of a fluid consisting of a huge number of discrete lines, like the spectrum of an atom.
       It is only necessary that the intensities of lines be sufficiently high. However, if the resonator disk is only covered by a helium film,
       no lines will be observed. Indeed, by (\ref{e37}), the intensities of lines will be of the same order of magnitude as those for
       a thick layer of helium \cite{svh1,svh3}. But the line registered for such a layer consists of $\sim 10^5$ individual lines
       which are too weak to be observed separately.
       To resolve them, one needs to increase their intensity by
       4-5 orders of magnitudes. This is a task for future studies.

               \section{Quantization of the amplitude of the roton line}

                The experiment \cite{svh2} revealed one more unusual effect: as the power $\dot{Q}$ of a heat gun increases,
       the amplitude $A_R$ of the roton absorption peak decreases, and this occurs stepwise. This fact testifies to the ``quantization'' of the roton line amplitude.
        It was noted in \cite{svh2} that this effect can be related
        to the quantization of the azimuth velocity $\textbf{v}_s$ (around
       the resonator), but the nature of this connection is not clear yet. Since the growth of $A_R$ means a decrease in the
       number of c-rotons created by SHF-photons, we assume that each step of $A_R$
       means a decrease in the number of created c-rotons by some integer.

          In He II near the resonator, two competing processes occur:
          c-photons create c-rotons and \textit{vice versa}.
                              In this case, a c-roton
         can transit only in a c-photon with the same energy and the same $l$.
                  From the state $|N_{phot},N_{rot}\rangle
         $ with $N_{phot}$ c-photons and $N_{rot}$ c-rotons, the transition in two following states
         is possible: \textsl{i}) the state $|N_{phot}+1,N_{rot}-1\rangle $, if a c-roton creates a c-photon; \textsl{ii})
         the state $|N_{phot}-1,N_{rot}+1\rangle $, if a c-photon creates a
         c-roton.

     Since circular photons and rotons are bosons, we can associate
         the creation operators with them. Within the formalism of secondary quantization for bosons, we obtain
         the transition probability (\textsl{i})
         \begin{equation}
     w(rot\mapsto phot) =  G(N_{phot}+1)N_{rot}.
     \label{wrp-1} \end{equation}
     For (\textsl{ii}), we have
       \begin{equation}
     w(phot\mapsto rot) =  G(N_{rot}+1)N_{phot},
     \label{wrp-2} \end{equation}
        whence
         \begin{eqnarray}
     { \sss\triangle} w &\equiv & w(rot\mapsto phot)-w(phot\mapsto rot) = \nonumber \\
     &=& G(N_{rot}-N_{phot}),
     \label{w-w} \end{eqnarray}
     where $G = w(phot\mapsto rot)$ at  $N_{phot}=1$, $N_{rot}=0$. In other words, $G $ is the above-calculated probability (\ref{a57})
      of the c-photon $\rightarrow$ c-roton transition in the case where there is a single c-photon in the initial state of the system, and
     there are no c-rotons.
      If ${ \sss\triangle} w < 0$, then the roton line in the spectrum of an SHF signal
      is a line of absorption of photons; but if ${ \sss\triangle} w > 0,$ the roton line is a line of their
      emission (``maser'' effect). In the experiment,
      the absorption line becomes weaker, as the power of a heat gun increases and transits to the emission line at $\dot{Q}=\dot{Q}_c \approx 0.5\,\mbox{W}/\mbox{cm}^2$.
      We assume that these facts are related to the forced creation of
       c-rotons by a gun. Though all details of the mechanism are unclear up to now, we indicate
      several points.

      a) A gun is directed along a tangent to
      the disk, i.e., so that the transfer of an angular momentum should be
      maximum. The p-rotons have no angular momentum, but the c-rotons
      have ($L_{z}=\hbar l_{c}$). Therefore, their creation is accompanied by the transfer of both the energy
      and the angular  momentum. It was noticed in \cite{svh2} that a step of $\dot{Q}$
          corresponds to an increase in $v_s$ on the output of a gun by a value coinciding with the
         quantum $\hbar  /m_4 R_{d}$
          of a circular velocity of helium near the resonator, according to the formula $v_s^{\varphi}=n_c\hbar  /m_4R_{d}$.
         The maximum experimental number of steps was $n_c^{max} \sim 3\cdot
         10^5$. Such a quantization of the velocity can mean that a
         gun creates a vortex in the superfluid component near the disk, whose axis coincides with the disk axis, and each step corresponds to the increase of the
         circulation by 1. On the other hand,
         the action of the operator of azimuth momentum $-\frac{i\hbar\partial}{\rho_j\partial\varphi_j}$
                on the factor $e^{il\varphi_j}$ of the WF of a
          c-roton gives $\hbar l /\rho_j$. Dividing it by
          the $^4$He atom mass, we obtain the azimuth
          velocity $v_{\varphi}=\hbar l /m_4\rho_j$.
           Thus, a c-roton induces the quantized rotation of helium atoms around the resonator with the velocity
           $v_{\varphi}=\hbar l /m_4\rho$, like a vortex. But this rotation is accompanied by simultaneous oscillations.
            In turn, an increase in the circular velocity of helium $v_{\varphi}$ by an external factor must
            stimulate the creation of c-phonons (including c-rotons).
            Apparently, a gun creates a macroscopic vortex in helium
            (or a superposition of vortices) and many plane
            and circular phonons and rotons, so that the ensemble of
            quasiparticles near the disk is quite complicated. Moreover, some processes (e.g., decays) are allowed for
            c-phonons and forbidden for p-phonons.

      b)   The results of experiments with a gun testify unambiguously that
              an EM wave creates namely c-, rather than p-rotons. If the latter would be created,
              then a decrease in the amplitude of the roton absorption line would be related, according to (\ref{w-w}), only to that
               a gun increases their number. But it increases also at a simple increase in the temperature (without
               switching-on a gun); however, the measurements show that this does not cause a weakening of the roton line and its transition into an emission line.

         c)  In a disk resonator, ``left'' (L) and
         ``right'' (R) c-photons differing by the sign of $l$ are created. A waveguide captures
         the total signal from the resonator, i.e., the summary field
         of L- and R-waves. In this case, the amplitude of one of these waves in the resonator is $\sim 100$ times greater than another one \cite{rub3}.
           However, a gun creates c-rotons with only a single polarization,
           L- or R-, which depends on the position of a gun. Since an L c-photon can
            induce only an L c-roton (the same is true, respectively, for right ones),
            Eq. (\ref{w-w}) should be written separately for L- and
           R-quanta. Then we obtain that the summary absorption line
           disappears if the relation $N_{rot} =N_{phot}$ holds true for the dominant photon mode. A gun is able to weaken the line if the signs of $l$ for a c-roton and
           the dominant c-photon coincide and cannot weaken if the signs do not coincide.
           Thus, let a gun weaken the line at a given
           configuration of the EM field. But if a gun is reoriented
           so that it twists He II in the
           opposite direction, it must stop to weaken
           the line.
           It seems to us that this assertion can be easily verified in experiments.

                \section{Stark effect}
             Finally, we mention the observation of the linear Stark effect in helium-II ---
      the roton absorption line splits into two lines
      in a constant electric field $\textbf{E}_0$ directed in the disk plane. The distance between them increases
      $\sim E_0$ \cite{svh4}. In \cite{mel2,min}, the authors advanced the idea of the relation of the effect to a possible quadrupole or
     instantaneous  dipole, respectively,  moment of a p-roton.
       We agree that a p-roton as a wave packet can possess an
       instantaneous DM $\textbf{d}_{r}$ with fluctuating (or pulsing)
value and direction. This DM is due to, in particular,
      the interaction of the roton with neighboring
      quasiparticles located nonuniformly. Such a DM induces an
      addition $\sim \textbf{d}_{r}\textbf{E}_0$ to the roton
      energy. Since the projection of $\textbf{d}_{r}$ on $\textbf{E}_0$
       takes the values in the continuous band
$[-\bar{d}^{z}_{r}E_{0},\bar{d}^{z}_{r}E_{0}]$,
        the mentioned addition transforms the roton level to the
         band. However, experiments demonstrate the splitting of the
        line into two ones, rather than a single band. Thus, the
        instantaneous  DM of a roton cannot explain the observed line
splitting.

        According to quantum mechanics, the effect can be explained
        if the roton energy level possesses at least a twofold degeneration
 which is taken off by the field $\textbf{E}_0$.
           In our opinion, the effect is determined by the existence of right ($l=l_{rot}$) and
      left ($l=-l_{rot}$) c-rotons with the same energy,
          whose superposition is described by the wave function $
      C_1\psi^{cir}_{rot}(l)+C_2\psi^{cir}_{rot}(-l).$
       Since a perturbing potential is obviously proportional to the applied field
       $\textbf{E}_0$, we obtain the splitting of a twofold degenerate energy level which is proportional to $E_0.$
       It can be evaluated by the well-known formulas of perturbation theory for a degenerate state. The field $\textbf{E}_0$
       induces the polarization of the dielectric
       resonator, which is directed along $\textbf{E}_0,$ and takes off the degeneration, by breaking the circular symmetry.
        But the multiple degeneration by $k_z$ remains
        (several tens of small $k_z$ significantly contribute to the line).
         So, if the field $\textbf{E}_0$ will be directed along the $Z$ axis, then the line must be split into two lines due to the removal
        of the degeneration relative to a change in the sign of $k_z$.

       At the switching-on of a heat gun, we may expect the following. In the
       field $\textbf{E}_0,$ the eigenfunctions of the Hamiltonian of
       He II are not the $R$-  and $L$-functions,
       $\psi^{cir}_{rot}(l)$ and $\psi^{cir}_{rot}(-l)$, but their superpositions
       $ \psi^{cir}_{rot}(l)\pm \psi^{cir}_{rot}(-l)$ characteristic of
       a twofold degenerate level \cite{vac}. The gun
       oriented along a tangent to the disk creates the quantized
       rotation of helium in a \textit{single} direction and increases the number of c-rotons with
       one ($R$ or $L$) polarization. It is obvious that, in this case, the gun cannot excite the states
       $ \psi^{cir}_{rot}(l)\pm \psi^{cir}_{rot}(-l)$ ($R\pm
       L$-superposition of two c-rotons with the counter rotation). Therefore, if the field
        $\textbf{E}_0$ was already switched-on (earlier than the gun), the gun will not decrease the peaks of the split roton
        line and will not induce the maser effect. It would be of interest to verify this prediction in experiments in similar fashion.

          \section{Conclusion}

       It is seen from the above-presented analysis that the experiment \cite{svh1,svh3}
       has revealed the existence of particular excitations in He II
       --- circular rotons which are azimuth sound waves.
            In the present work, we have approximately calculated the probability of the creation of a circular roton by the EM field of the resonator
       and, on its basis, have evaluated the width of the absorption line at the roton
       frequency. The theoretical line is close to the experimental one by width but differs by shape.

      We have also advanced the assumption that the splitting of the line into two ones in a
       constant electric field is caused by the presence of right and left c-rotons
       and, respectively, by the twofold degeneration of the energy level of a circular
       roton.

   \vskip3mm
       The authors are  grateful to V.\,N. Derkach, E.\,Ya. Rudavskii, and A.\,S. Rybalko
       for numerous discussions of the experiment and to Yu.\,V. Shtanov for valuable remarks and advices.


\begin{thebibliography}{200}
  \bibitem {svh1} A.~Rybalko, S.~Rubets, E.~Rudavskii, V.~Tikhiy, S.~Tarapov,
    R.~Golovashchenko, and V.~Derkach, Phys. Rev. B \textbf{76},
      140503(R) (2007).
   \bibitem {svh3} A.S.~Rybalko, S.P.~Rubets, E.Ya.~Rudavskii, V.A.~Tikhiy, Yu.M.~Poluektov,
   R.V.~Golovashchenko,  V.N.~Derkach, S.I.~Tarapov, and O.V.~Usatenko
         Fiz. Nizk. Temp. \textbf{35}, 1073 (2009)
    [Low Temp. Phys. \textbf{35}, 837 (2009)].
   \bibitem {I} V.M.~Loktev and M.D.~Tomchenko, Ukr. J. Phys. \textbf{55}, 901 (2010)
      [www.ujp.bitp.kiev.ua/files/file/papers/ 55/8/550807p.pdf].
   \bibitem {II} V.M.~Loktev and M.D.~Tomchenko, Phys. Rev. B \textbf{82}, 172501 (2010).
   \bibitem {land3} L.D.~Landau, E.M.~Lifshitz, {\it Quantum Mechanics.
Non-Relativistic Theory} (Pergamon, New York, 1980).
    \bibitem {land4} V.B.~Berestetskii, E.M.~Lifshitz, and L.P.~Pitaevskii, {\it Relativistic Quantum Theory}
         (Pergamon Press, Oxford, 1982).
   \bibitem {wb} W. Byers Brown and D.M. Whisnant, Mol. Phys. \textbf{25}, 1385, \textbf{26}, 1105 (1973).
   \bibitem {lt2} V. M. Loktev and M. D. Tomchenko, J. Phys. B: At. Mol. Opt. Phys. \textbf{44}, 035006
 (2011);  Dop. Nats. Akad. Nauk Ukr. N 5, 76 (2010) (in Russian).
     \bibitem {pol2D} M.D. Tomchenko,  arXiv:cond-mat/1003.4389  (2010).
         \bibitem {y} E. Janke, F. Emde, F. L\"{o}sch,  \textit{Tafeln H\"{o}herer Funktionen} (Teubner, Stuttgart, 1960).
     \bibitem {rub3} A.S. Rybalko, private communication.
      \bibitem {and} M.R. Gibbs, K.H. Andersen,  W.G. Stirling, H.
        Schober, J.~Phys. Cond. Mat. \textbf{11}, 603 (1999).
     \bibitem {kris1}  A.S. Davydov, \textsl{Theory of Molecular Excitons}
         (Plenum, New York, 1971).
   \bibitem {kris2}   D. Pines, \textsl{Elementary Excitations in Solids}
         (Benjamin, New York, 1963).
    \bibitem {lt1} V.M. Loktev and M.D. Tomchenko, Fiz. Nizk. Temp. \textbf{34}, 337 (2008)
    [Low Temp. Phys. \textbf{34}, 262 (2008)].
    \bibitem {svh2}  A.S. Rybalko, S.P. Rubets, E.Ya. Rudavskii, V.A.~Tikhiy, R.V.~Golovashchenko,  V.N.~Derkach, S.I.~Tarapov,
    Fiz. Nizk. Temp. \textbf{34}, 326 (2008)
    [Low Temp. Phys. \textbf{34}, 254 (2008)].
     \bibitem {svh4} A.S. Rybalko, S.P. Rubets, E.Ya. Rudavskii \textit{et al.},
         arXiv:cond-mat/0807.4810 (2008).
\bibitem {mel2}  L.A. Melnikovsky, arXiv:cond-mat/0808.1188 (2008).
\bibitem {min}  V.P. Mineev, Pis'ma v Zh. Eksp. Teor. Fiz. \textbf{90}, 866
(2009)    [JETP Lett. \textbf{90}, 866 (2009)].
\bibitem {vac} I.O. Vakarchuk, \textsl{Quantum Mechanics}
   (L'viv National University, L'viv, 2004) (in Ukrainian).

\end{thebibliography}
               \end{document}